\documentclass[12pt,cite,epsf,epsfig]{article}
\usepackage{epsfig}

\begin{document}

\author{S. Dev\thanks{Electronic address: dev5703@yahoo.com} ,
 Surender Verma\thanks{Electronic address: s\_7verma@yahoo.co.in} , Shivani
Gupta\thanks{Electronic address: shiroberts\_1980@yahoo.co.in} }

\title{Phenomenological Analysis of Hybrid Textures of Neutrinos}
\date{\textit{Department of Physics, Himachal Pradesh University, Shimla 171005, INDIA.}\\
\smallskip}

\maketitle
\begin{abstract}
We present a comprehensive phenomenological analysis of the
allowed hybrid textures of neutrinos. Out of a total of sixty
hybrid textures with one equality between the elements of neutrino
mass matrix and one texture zero only twenty three are found to be
viable at 99\% C.L. whereas the earlier analysis found fifty four
to be viable. We examine the phenomenological implications of the
allowed hybrid textures including Majorana type CP-violating
phases, 1-3 mixing angle and Dirac type CP-violating phase,
$\delta$. We, also, obtain lower bound on effective Majorana mass
for all the allowed hybrid textures.
\end {abstract}

\section{Introduction}

In the Standard Model (SM), fermions acquire masses via
spontaneous breakdown of SU(2) gauge symmetry. However, the values
of fermion masses and the observed hierarchical fermion spectra
are not understood within the SM. This results in thirteen free
parameters in the SM which includes three charged lepton masses,
six quark masses and the four parameters of the CKM matrix. The
symmetries of the SM do not allow non-zero neutrino masses through
renormalizable Yukawa couplings. However, non-zero neutrino masses
can be introduced via non-renormalizable higher dimensional
operators presumably having their origin in physics beyond the SM.
Texture zeros, flavor symmetries, radiative mechanisms and the
see-saw mechanisms are some of the widely discussed mechanisms for
fermion mass generation. These mechanisms, most often, complement
and reinforce each other. In the ongoing decade, significant
advances have been made in understanding these mechanisms. In
particular texture zeros and flavor symmetries have provided
quantitative relationships between flavor mixing angles and the
quark/lepton mass ratios. It has, now been realized that the
``See-Saw GUT'' scenario, on its own, cannot provide a complete
understanding of the flavor structure of the quark and lepton mass
matrices and new physics seems to be essential perhaps in the form
of new symmetries mainly in the lepton sector. Moreover, a unified
description of flavor physics and CP-violation in the quark and
lepton sectors is absolutely necessary. This can be achieved by
constructing a low energy effective theory with the SM and some
discrete non-Abelian family symmetry and, subsequently, embedding
this theory into Grand Unified Theory (GUT) models like
SO(10)\cite{1}. For this reason, the discrete symmetry will have
to be a subgroup of SO(3) or SU(3)\cite{2}. The search for an
adequate discrete symmetry has mainly focussed on the minimal
subgroups of these groups with at least one singlet and one
doublet irreducible representation to accommodate the fermions
belonging to each generation. One such subgroup is the quaternion
group $Q_8$\cite{3} which not only accommodates the three
generations of fermions but also explains the rather large
difference between values of 2-3 mixings in the quark and lepton
sectors. Quaternion symmetry like some other discrete symmetries
leads to nontrivial relationships amongst the non-zero mass matrix
elements which underscores the inadequacy of texture zero
analyses\cite{4,5,6,7,8,9,10,11} in isolation. Such textures which have
equalities between different elements alongwith the vanishing of
some elements of the mass matrix have been referred to as hybrid
textures in the literature. Frigerio and Smirnov\cite{12}
presented a comprehensive analysis of the hybrid textures
alongwith other possibilities for the neutrino mass matrix. The
detailed numerical analysis of sixty possible hybrid textures with
one equality amongst mass matrix elements and one texture zero was
presented by Kaneko \textit{et al.} \cite{13} who found fifty four
out of sixty to be phenomenologically viable. Hybrid textures
have, also, been discussed more recently\cite{14} though in a
somewhat different context.

In the present work, we present a detailed and improved numerical
analysis of the hybrid textures found to be viable in the earlier
numerical analysis by Kaneko \textit{et al.} \cite{13}. Our analysis
shows that only twenty three out of a total of sixty possible
hybrid textures are phenomenologically viable. We examine the
phenomenological implications of the viable hybrid textures. These
hybrid textures have different hierarchical spectra and have
different phenomenological consequences. We find the values of
Majorana type CP-violating phases for which these textures are
realized. We, also, calculate the effective Majorana mass and
Jarlskog rephasing invariant $J_{CP}$ for all viable hybrid
textures of neutrinos. In addition, we present correlation plots
between different parameters of the hybrid textures of neutrinos
for $3{\sigma}$ allowed ranges of the known parameters. It is
found that positive results from neutrinoless double beta decay
experiments will rule out three more hybrid textures with one
equality between the matrix elements and one zero. Thus, the
investigation of Majorana type CP-violating phases will be
significant.

\section{Hybrid Textures}

The neutrino mass matrix, $M_{\nu}$, can be parameterized in terms
of three neutrino mass eigenvalues ($m_{1}$, $m_{2}$, $m_{3}$),
three neutrino mixing angles ($\theta_{12}$, $\theta_{23}$,
$\theta_{13}$) and one Dirac type CP-violating phase, $\delta$. If
neutrinos are Majorana particles then there are two additional
CP-violating phases $\alpha$, $\beta$ in the neutrino mixing
matrix. Thus, massive Majorana neutrinos increase the number of
free parameters of the SM from thirteen to twenty two. In the
charged lepton basis the complex symmetric mass matrix $M_{\nu}$
can be diagonalized by a complex unitary matrix $V$:
\begin{equation}
M_{\nu}=VM_{\nu}^{diag}V^{T}
\end{equation}
where $M_{\nu}^{diag}=Diag \{m_1,m_2,m_3\}$ is the diagonal
neutrino mass matrix. The neutrino mixing matrix $V$\cite{15} can
be written as
\begin{equation}
V\equiv U P=\left(
\begin{array}{ccc}
c_{12}c_{13} & s_{12}c_{13} & s_{13}e^{-i\delta} \\
-s_{12}c_{23}-c_{12}s_{23}s_{13}e^{i\delta} &
c_{12}c_{23}-s_{12}s_{23}s_{13}e^{i\delta} & s_{23}c_{13} \\
s_{12}s_{23}-c_{12}c_{23}s_{13}e^{i\delta} &
-c_{12}s_{23}-s_{12}c_{23}s_{13}e^{i\delta} & c_{23}c_{13}
\end{array}
\right)\left(
\begin{array}{ccc}
1 & 0 & 0 \\ 0 & e^{i\alpha} & 0 \\ 0 & 0 & e^{i(\beta+\delta)}
\end{array}
\right),
\end{equation}
where $s_{ij}=\sin\theta_{ij}$ and $c_{ij}=\cos\theta_{ij}$. The
matrix $V$ is called the neutrino mixing matrix or PMNS matrix.
The matrix $U$ is the lepton analogue of the CKM quark mixing
matrix and the phase matrix $P$ contains the two Majorana phases.
Therefore the neutrino mass matrix can be written as
\begin{equation}
M_{\nu}=U P M_{\nu}^{diag}P^{T}U^{T}.
\end{equation}
The elements of the neutrino mass matrix can be calculated from
Eq. (3). In hybrid textures of neutrinos, we have one equality of
matrix elements
 and one zero.i.e.
 \begin{center}
 \begin{eqnarray}
M_{ab}=0,\\ \nonumber M_{pq}= M_{rs}.
\end{eqnarray}
 \end{center}
 These two conditions yield two complex equations as follows
 \begin{equation}
m_{1}U_{a1}U_{b1}+m_{2}U_{a2}U_{b2}e^{2i\alpha
}+m_{3}U_{a3}U_{b3}e^{2i(\beta +\delta )}=0
\end{equation}
and
\begin{equation}
m_{1}(U_{p1}U_{q1}-U_{r1}U_{s1})+m_{2}(U_{p2}U_{q2}-U_{r2}U_{s2})e^{2i\alpha
}+m_{3}(U_{p3}U_{q3}-U_{r3}U_{s3})e^{2i(\beta +\delta )}=0
\end{equation}
where $U$ has been defined in Eq. (2). These two complex equations
involve nine physical parameters $m_{1}$, $m_{2}$, $m_{3}$,
$\theta _{12}$, $\theta _{23}$, $\theta _{13}$ and three
CP-violating phases $\alpha $, $\beta $ and $\delta $. The masses
$m_{2}$ and $m_{3}$ can be calculated from the mass-squared
differences $\Delta m_{12}^{2}$ and $\Delta m_{23}^{2}$ using the
relations
\begin{equation}
m_{2}=\sqrt{m_{1}^{2}+\Delta m_{12}^{2}}
\end{equation}
and
\begin{equation}
m_{3}=\sqrt{m_{2}^{2}+\Delta m_{23}^{2}}.
\end{equation}
Thus, we have two complex equations relating five unknown
parameters viz. $m_1$, $\theta_{13}$, $\alpha$, $\beta$ and
$\delta$. Therefore, if one out of these five parameters is
assumed, other four parameters can be predicted. Solving Eqs. (5)
and (6) simultaneously, we obtain
\begin{equation}
\frac{m_{1}}{m_{3}}e^{-2i\beta }=\frac{%
U_{a2}U_{b2}(U_{p3}U_{q3}-U_{r3}U_{s3})-U_{a3}U_{b3}(U_{p2}U_{q2}-U_{r2}U_{s2})}{
U_{a1}U_{b1}(U_{p2}U_{q2}-U_{r2}U_{s2})-U_{a2}U_{b2}(U_{p1}U_{q1}-U_{r1}U_{s1})} e^{2i\delta }
\end{equation}
and
\begin{equation}
\frac{m_{1}}{m_{2}}e^{-2i\alpha }=\frac{%
U_{a3}U_{b3}(U_{p2}U_{q2}-U_{r2}U_{s2})-U_{a2}U_{b2}(U_{p3}U_{q3}-U_{r3}U_{s3})}{
U_{a1}U_{b1}(U_{p3}U_{q3}-U_{r3}U_{s3})-U_{a3}U_{b3}(U_{p1}U_{q1}-U_{r1}U_{s1})}.
\end{equation}
Using Eqs. (9) and (10), the two mass ratios
$\left(\frac{m_1}{m_2},\frac{m_1}{m_3}\right)$ and the two
Majorana phases ($\alpha $, $\beta $) can be written as
\begin{equation}
\frac{m_1}{m_3}=\left| \frac{
U_{a2}U_{b2}(U_{p3}U_{q3}-U_{r3}U_{s3})-U_{a3}U_{b3}(U_{p2}U_{q2}-U_{r2}U_{s2})}{
U_{a1}U_{b1}(U_{p2}U_{q2}-U_{r2}U_{s2})-U_{a2}U_{b2}(U_{p1}U_{q1}-U_{r1}U_{s1})}\right|,
\end{equation}
\begin{equation}
\frac{m_1}{m_2}=\left| \frac{
U_{a3}U_{b3}(U_{p2}U_{q2}-U_{r2}U_{s2})-U_{a2}U_{b2}(U_{p3}U_{q3}-U_{r3}U_{s3})}{
U_{a1}U_{b1}(U_{p3}U_{q3}-U_{r3}U_{s3})-U_{a3}U_{b3}(U_{p1}U_{q1}-U_{r1}U_{s1})}\right|,
\end{equation}

\begin{equation}
\alpha =-\frac{1}{2}arg\left( \frac{%
U_{a3}U_{b3}(U_{p2}U_{q2}-U_{r2}U_{s2})-U_{a2}U_{b2}(U_{p3}U_{q3}-U_{r3}U_{s3})}{
U_{a1}U_{b1}(U_{p3}U_{q3}-U_{r3}U_{s3})-U_{a3}U_{b3}(U_{p1}U_{q1}-U_{r1}U_{s1})}\right)
\end{equation}
and
\begin{equation}
\beta =-\frac{1}{2}arg\left( \frac{
U_{a2}U_{b2}(U_{p3}U_{q3}-U_{r3}U_{s3})-U_{a3}U_{b3}(U_{p2}U_{q2}-U_{r2}U_{s2})}{
U_{a1}U_{b1}(U_{p2}U_{q2}-U_{r2}U_{s2})-U_{a2}U_{b2}(U_{p1}U_{q1}-U_{r1}U_{s1})}\right)-\delta.
\end{equation}
Both the Majorana phases $\alpha$ and $\beta$ in Eqs. (13) and
(14) are functions of $\theta_{13}$ and $\delta$ whereas the mass
ratios $\left(\frac{m_1}{m_2},\frac{m_1}{m_3}\right)$ are
functions of $\theta_{13}$ and $\delta$ since $\theta_{12}$ ,
$\theta_{23}$, $\Delta m^2_{12}$ and $\Delta m^2_{23}$ are known
experimentally. The values of mass ratios
$\left(\frac{m_1}{m_2},\frac{m_1}{m_3}\right)$, from Eqs. (11) and
(12) can be used to calculate $m_1$. The mass eigenvalue $m_1$ can
be written in terms of two mass squared differences which are
known from experiments.
\begin{equation}
m_{1}=\frac{m_1}{m_2} \sqrt{\frac{ \Delta
m_{12}^{2}}{1-\left(\frac{m_1}{m_2}\right) ^{2}}}
\end{equation}
and
\begin{equation}
m_{1}=\frac{m_1}{m_3} \sqrt{\frac{\Delta m_{12}^{2}+ \Delta
m_{23}^{2}}{ 1-\left(\frac{m_1}{m_3}\right)^{2}}}.
\end{equation}
The two values of $m_{1}$ obtained above from the two mass ratios
 must be equal to within the
errors of the mass squared differences. Using the experimental
inputs of the two mass squared
 differences and the two mixing angles we can constrain the ($\theta_{13}$, $\delta$) plane.
 The experimental constraints on neutrino parameters
with $1\sigma $\cite{16} errors are given below:
\begin{eqnarray}
\Delta m_{12}^{2} &=&7.67_{-0.21}^{+0.22}\times
10^{-5}eV^{2}, \nonumber \\ \Delta m_{23}^{2} &=&\pm
2.37_{-0.15}^{+0.15}\times 10^{-3}eV^{2},  \nonumber \\
\theta_{12}& =&34.5_{-1.4}^{+1.4}, \nonumber \\ \theta_{23}
&=&42.3_{-3.3}^{+5.1}.  \nonumber
\end{eqnarray}
Only an upper bound is known on the mixing angle $\theta_{13}$
from the CHOOZ experiment.
 We vary the oscillation
parameters $\delta$ and $\theta_{13}$ within their full physical
ranges with uniform distributions. We, also, calculate the
effective Majorana mass, which is given as
\begin{equation}
M_{ee}= |m_1c_{12}^2c_{13}^2+m_2s_{12}^2c_{13}^2 e^{2i\alpha}+m_3s_{13}^2e^{2i\beta}|
\end{equation}
 and Jarlskog rephasing
invariant quantity\cite{17}
\begin{equation}
J_{CP}=s_{12}s_{23}s_{13}c_{12}c_{23}c_{13}^2 \sin \delta
\end{equation}
within the allowed parameter space for all allowed hybrid textures. In our numerical analysis we generate as many as $10^8$ data points ($10^5$ data points were generated in the previous analysis\cite{13}) thus, making our numerical analysis more reliable.
Earlier analysis of hybrid textures\cite{13} used the ratio of two known
mass-squared differences
\begin{equation}
R_{\nu}\equiv \frac{\Delta m^2_{12}}{\Delta m^2_{23}}
=\frac{1-(\frac{m_1}{m_2})^2}{\left|1-\frac{(\frac{m_1}{m_2})^2}
{(\frac{m_1}{m_3})^2}\right|}
\end{equation}
to constrain the other neutrino parameters. However, our analysis
makes direct use of the two mass-squared differences and is more
constraining. Earlier analysis does not use the full experimental
information currently available with us in the form of two
mass-squared differences. Moreover, since $R_{\nu}$ is a function
of mass ratios, it does not depend upon the absolute neutrino mass
scale. It is obvious that the two mass ratios $\frac{m_1}{m_2}$
and $\frac{m_1}{m_3}$ may yield the experimentally allowed values
of $R_{\nu}$ for mutually inconsistent values of $m_1$. However,
our analysis selects only those mass ratios for which the values
of $m_1$ obtained from Eqs. (15) and (16) are identical to within
the errors of the mass squared differences. Moreover, the
definition of $R_{\nu}$ used in many earlier analysis does not
make use of the knowledge of solar mass hierarchy to constrain the
neutrino parameter space. Our numerical analysis restricts the
number of allowed hybrid textures to twenty three at 99\% C.L. as
enumerated in Table 1.

\section{Results And Discussions}

All the allowed hybrid textures of neutrinos give hierarchical
spectrum of neutrino masses. There are ten textures with clear
normal hierarchy, two with inverted hierarchy, and one with all
hierarchies. The remaining ten textures give normal and
quasi-degenerate spectrum as given in Table 2.

Table 3. depicts the lower bound on $\theta_{13}$ for viable
hybrid textures. The measurement of $\theta_{13}$ is the main goal
of future experiments. The proposed experiments such as Double
CHOOZ plan to explore $\sin^22\theta_{13}$ down to 0.06 in phase I
(0.03 in phase II)\cite{18}. Daya Bay has a higher sensitivity and
plans to observe $\sin^22\theta_{13}$ down to 0.01\cite{19}. Thus
some of these textures will face stringent experimental scrutiny
in the immediate future. We have, also, calculated the effective
Majorana mass which is the absolute value of $M_{ee}$ element of
neutrino mass matrix. The neutrinoless double beta decay which is
controlled by  effective Majorana mass is forbidden for texture
$C1$, $C2$ and $C3$ as $M_{ee}=0$ for these textures. Table 4.
gives the lower bound on $M_{ee}$ for the remaining hybrid
textures of neutrinos.

The analysis of $M_{ee}$ will be significant as many neutrinoless
double beta decay experiments will give the predictions on this
parameter. A most stringent constraint on the value of $M_{ee}$
was obtained in the $^{76}Ge$
 Heidelberg-Moscow experiment\cite{20} $|M_{ee}|< 0.35$eV.
 There are large number of projects such as SuperNEMO\cite{21},
 CUORE\cite{22}, CUORICINO\cite{22} and  GERDA\cite{23} which
 aim to achieve a sensitivity below 0.01eV to $M_{ee}$.
 Forthcoming experiment SuperNEMO, in particular, will explore
 $M_{ee}$ $<$ 0.05eV\cite{24}. The Jarlskog rephasing invariant
 quantity $J_{CP}$ varies in the range  (-0.05-0.05) for most of the allowed hybrid textures.
We get a clear normal hierarchical mass spectrum for ten hybrid
texture structures (Table 2). The Dirac type CP-violating phase
$\delta$ is constrained to the range $90^{o}$ - $270^{o}$ for $A1$
as seen in Fig.1(a). It can be seen from Fig.1(a) and Fig.1(c)
that there exists a clear lower bound on $\theta_{13}$ and
effective Majorana mass $M_{ee}$ . However, this range of $\delta$
is disallowed for $B2$. For $B1$ the Majorana type CP-violating
phase $\alpha$ is constrained to a value of  $0^{o}$ or $180^{o}$
as seen in Fig.1(d). However, its range varies from $50^{o}$ to
$130^{o}$ for $C1$, $C2$ and $C3$ and this range of $\alpha$ is
disallowed for $A1$ and $B2$. The Majorana phase $\beta$ is
unconstrained i.e. whole range of $\beta$ is allowed for all
textures with normal hierarchy. Maximal value of $\theta_{23}$ is
disallowed for $A3$, $A5$, $B4$, $B6$, $C1$, $C2$ and $C3$.
Fig.1(b) depicts Jarlskog rephasing invariant quantity $J_{CP}$ as
a function of $\delta$ for $B1$. The allowed range of $J_{CP}$ for
$B1$ is constrained to be (-0.018-0.018).

The other ten textures give quasidegenerate spectrum in addition
to normal hierarchy. We get the lower bound on effective Majorana
mass (Fig.2(a)) and the reactor mixing angle. Many forthcoming
neutrino oscillation experiments aim to measure/constrain these
two parameters. The Dirac type CP-violating phase $\delta$ is
constrained for most of these hybrid texture structures as seen in
Fig.2(a) for $B3$ . Fig.2(b) shows the mass spectrum for $B3$. The
correlation plot shows both normal and quasidegenerate behaviour.
 Dirac type CP-violating $\delta$ lies in the range $90^{o}$ or
 $270^{o}$ for $B3$, $B5$ while this range is disallowed for $A2$,
 $A4$, $D1$, $D3$, $D4$, $D5$, $E3$ and $E4$. The
 Majorana-type CP-violating phase, $\alpha$, takes a single value of
 $0^{o}$ or $180^{o}$ for $B3$, $B5$, $D1$, $D3$, $D4$ and $E4$.
 We find some hybrid textures having identical predictions for all
 parameters except the 2-3 mixing angle i.e. $\theta_{23}$ is above
 maximal for one and below maximal for the other set. Such textures are
 given in Table.5. There are some projects like Tokoi- to- Kamioka-
 Korea (T2KK) which intend to resolve the octant degeneracy of
 $\theta_{23}$ (i.e. $\theta_{23}<45^{o}  $ or $\theta_{23}>45^{o}$)\cite{25}.
 The results from future experiments will decide the fate of such texture structures.
A comparative study of some predictions of the present work and
the earlier analysis\cite{13} is presented in Table.6.

Two hybrid texture structures viz. $D2$, $E2$ have inverted
hierarchical mass spectrum. We find identical predictions for all
the neutrino parameters for these two textures except for the
Majorana type CP-violating phase $\alpha$ as given in Fig.3.,
which is restricted to a range of $82^{o}$ to $96^{o}$.

For the remaining texture $E1$ all mass hierarchies are possible
as shown in correlation plot 4. The Majorana phase $\alpha$ is
constrained to a value of $0^{o}$ or $180^{o}$, while no
constraint is obtained for the other Majorana phase $\beta$ in
this hybrid texture.

\section{Conclusions}
In conclusion, we find that only twenty three hybrid textures with
one equality between mass matrix elements and one texture zero are
allowed at 99\% C.L. by our analysis. We present systematic and
detailed numerical analysis for twenty three allowed hybrid
textures of neutrinos. These hybrid textures have different
hierarchical spectra and, thus, have different phenomenological
implications. Ten out of twenty three allowed hybrid textures give
normal hierarchical mass spectra, two imply inverted hierarchy and
one texture has all hierarchies. The remaining ten hybrid textures
give quasidegenerate spectra in addition to normal hierarchy.
Predictions for 1-3 mixing angle and Dirac type CP-violating
phase, $\delta$, are given for these textures. These two
parameters are expected to be measured in the forthcoming neutrino
oscillation experiments. We, also, obtained the lower bound on
effective Majorana mass for all the allowed hybrid textures. The
possible measurement of effective Majorana mass in neutrinoless
double $\beta$ decay experiments will provide an additional
constraint on the remaining three neutrino parameters i.e. the
neutrino mass scale and two Majorana type CP-violating phases. The
observation of correlations between various neutrino parameters
like $\theta_{13}$, $\theta_{23}$, $\delta$, $M_{ee}$ etc will
confirm/reject hybrid textures with a texture zero and an
additional equality among the matrix elements.
\newpage

\textbf{\textit{\Large{Acknowledgements}}}

The research work of S. D. is supported by the University Grants
Commission, Government of India \textit{vide} Grant No. 34-32/2008
(SR). S. G. and S. V. acknowledges the financial support provided
by Council for Scientific and Industrial Research (CSIR) and
University Grants Commission (UGC), Government of India,
respectively.

\newpage

\begin{table}[h]
\begin{center}
\begin{tabular}{||c|c|c|c|c|c||}
 \hline
   &A & B  & C & D & E  \\
 \hline\hline
 1 &  $\left(
\begin{array}{ccc}
X & 0 & e \\  & b & X \\ &  & c
\end{array}
\right)$&$\left(
\begin{array}{ccc}
X & X & 0 \\  & b & f \\ &  & c
\end{array}
\right)$  & $\left(
\begin{array}{ccc}
0 & d & e \\  & X & f \\ &  & X
\end{array}
\right)$  &$\left(
\begin{array}{ccc}
a & X & X \\  & 0 & f \\ &  & c
\end{array}
\right)$  & $\left(
\begin{array}{ccc}
a & X & X \\  & b & f \\ &  & 0
\end{array}
\right)$ \\
 \hline
2  & $\left(
\begin{array}{ccc}
a & 0 & X \\  &X & f \\ &  & c
\end{array}
\right)$& $\left(
\begin{array}{ccc}
X & d & 0 \\  & b & X \\ &  & c
\end{array}
\right)$  &  $\left(
\begin{array}{ccc}
0 & d & e \\  & X & X \\ &  & c
\end{array}
\right)$ & $\left(
\begin{array}{ccc}
X & d & e \\  & 0 & X \\ &  & c
\end{array}
\right)$ & $\left(
\begin{array}{ccc}
X & d & e \\  & b & X \\ &  & 0
\end{array}
\right)$\\
\hline 3  & $\left(
\begin{array}{ccc}
a & 0 & e \\  & X & X \\ &  & c
\end{array}
\right)$ & $\left(
\begin{array}{ccc}
a & X & 0 \\  & X & f \\ &  & c
\end{array}
\right)$&$\left(
\begin{array}{ccc}
0 & d & e \\  & b & X \\ &  & X
\end{array}
\right)$& $\left(
\begin{array}{ccc}
a & X & e \\  & 0 & f \\ &  &X
\end{array}
\right)$ & $\left(
\begin{array}{ccc}
a& X & e \\  & X & f \\ &  & 0
\end{array}
\right)$\\
\hline 4  &$\left(
\begin{array}{ccc}
a & 0 & X \\  & b & f \\ &  & X
\end{array}
\right)$ & $\left(
\begin{array}{ccc}
a & d & 0 \\  & X & X \\ &  & c
\end{array}
\right)$  & -& $\left(
\begin{array}{ccc}
a & d & X \\  & 0 & f \\ &  & X
\end{array}
\right)$ & $\left(
\begin{array}{ccc}
a & d & X \\  & X &f \\ &  & 0
\end{array}
\right)$ \\
\hline 5  & $\left(
\begin{array}{ccc}
a & 0 & e \\  & b & X \\ &  & X
\end{array}
\right)$ & $\left(
\begin{array}{ccc}
a & X & 0\\  & b & f \\ &  & X
\end{array}
\right)$ & - & $\left(
\begin{array}{ccc}
a & d & e \\  & 0 & X \\ &  & X
\end{array}
\right)$& -  \\
\hline 6  & - &$\left(
\begin{array}{ccc}
a & d & 0 \\  & b & X \\ &  & X
\end{array}
\right)$ & - & - & -     \\
\hline
\end{tabular}
\caption{Twenty three allowed hybrid textures. $X$ denote two
equal elements.}
\end{center}
\end{table}

\begin{table}
\begin{center}
\begin{tabular}{||c|c|c|c|c|c||}
 \hline
   &A & B  & C & D & E  \\
 \hline\hline
 1 &  NH  & NH  &  NH   & NH + QD  & NH+IH+QD \\
 \hline
2  &  NH+ QD &  NH  &  NH & IH & IH \\
\hline
3  & NH & NH + QD & NH & NH + QD & NH + QD \\
\hline
4  &NH + QD & NH  & -& NH + QD & NH + QD \\
\hline
5  & NH & NH + QD & - & NH + QD & -  \\
\hline
6  & - & NH  & - & - & -     \\
\hline
\end{tabular}
\caption{Mass patterns for the allowed hybrid textures.}
\end{center}
\end{table}
\begin{table}[tb]
\begin{center}
\begin{tabular}{||c|c||}
 \hline
 $\theta_{13}$  & Hybrid Textures   \\
 \hline\hline
 $> 0^{o}$ &     $A2$, $A4$, $B3$, $B5$, $B6$, $C1$, $D3$, $D4$, $E3$, $E4$       \\
 \hline
$\ge 1^{o}$  &     $A1$, $A3$, $A5$, $B2$, $E1$    \\
\hline
$\ge 2^{o}$  & $B1$, $B4$, $D1$, $D5$ \\
\hline

\end{tabular}
\caption{Prediced Lower Bounds on $\theta_{13}$.}
\end{center}
\end{table}
\begin{table}
\begin{center}
\begin{tabular}{||c|c||}
 \hline
 $M_{ee}(eV)$  & Hybrid Textures   \\
 \hline\hline
 $>$ 0 &     $A2$, $A3$, $A4$, $A5$, $B3$, $B4$, $B6$       \\
 \hline
$\ge 0.003eV$  &      $B1$     \\
\hline
$\ge 0.01eV$  & $A1$, $B2$, $D1$, $D2$, $D5$, $E1$, $E2$ \\
\hline
$\ge 0.02eV$  &   $D3$, $D4$, $E3$       \\
\hline
$\ge 0.03eV$  &  $E4$  \\
\hline
$\ge 0.05eV$  &   $B5$   \\
\hline
\end{tabular}
\caption{Predicted Lower Bounds on $M_{ee}$.}
\end{center}
\end{table}

\begin{table}
\begin{center}
\begin{tabular}{||c|c||}
 \hline
 Hybrid Textures  &  $\theta_{23}$  \\
 \hline\hline
$A2$  &    $<45^{o}$  \\
$A4$   &  $>45^{o}$ \\
 \hline
$B3$ &      $<45^{o}$  \\
$B5$ &    $>45^{o}$  \\
\hline
$D3$ &      $<45^{o}$   \\
$E3$ &    $>45^{o}$  \\
\hline
$D4$ &      $<45^{o}$   \\
$E4$ &    $>45^{o}$  \\
\hline
\end{tabular}
\caption{The hybrid textures in the same block give identical
predictions except $\theta_{23}$.}
\end{center}
\end{table}
\newpage
\begin{table}[h]
\begin{center}
\small
\begin{tabular}{||c|c|c|c|c||}
 \hline
Hybrid Texture & Hierarchy   & $\theta_{13}$ & $M_{ee}$ (eV) (Lower Bound)& $\theta_{23}$  \\
 \hline\hline
  A1  & NH & $>1^{o}$  & $\ge 0.01$&- \\
I I & NH + QD &- & 0.01-0.03& -\\
 \hline
   A2 & NH + QD & $>0^{o}$  &  $>0 $ & $<45^{o}$\\
K I & All &$\ge0^{o}$  &0.01-0.03& -\\
\hline
 A3  & NH & $>1^{o}$  &  $>0 $ &- \\
L I & All &- & 0.0001-0.0003& -\\
\hline
 A4 & NH + QD & $ >0^{o}$  & $>0 $ & $>45^{o}$ \\
N I & All  & $\ge0^{o}$& 0.001-0.003 & -\\
\hline
  A5 & NH &$ >1^{o}$   & $>0 $ & -\\
O I &All &- &0.0001-0.0003 &- \\
\hline
 B1 & NH & $ >2^{o}$  &$\ge 0.003$ & -\\
G II & NH &$ >2.2^{o}$ & 0.001-0.003& -\\
\hline
 B2  & NH & $ >1^{o}$  &$\ge 0.01$ &- \\
I II & NH+ QD & -&0.01-0.03 & -\\
\hline
  B3& NH+ QD & $ >0^{o}$  & $>0 $ & $<45^{o}$ \\
 J II& All &$\ge0^{o}$&0.001-0.003 &- \\
\hline
 B4 & NH &$ >2^{o}$     &$>0 $  &- \\
 L II& All&- & 0.0001-0.0003 &- \\
\hline
 B5 & NH+ QD &$ >0^{o}$     &$\ge 0.05 $  & $>45^{o}$\\
 M II & All &$\ge0^{o}$ & 0.01-0.03 &- \\
\hline
  B6& NH & $ >0^{o}$  &$>0 $  &- \\
 O II& All& -& 0.0001-0.0003&- \\
\hline
  C1& NH & $ \ge 0^{o}$  & 0  &- \\
 C IV& NH & -& 0  &- \\
\hline
  C2& NH & $ \ge 0^{o}$  & 0 &- \\
 L IV& NH &- & 0 & -\\
\hline
 C3 & NH & $ \ge 0^{o}$   & 0 &- \\
 O IV& NH & - & 0  &- \\
\hline
  D1& NH+ QD &$ >2^{o}$     &$\ge 0.01$ &- \\
 D V& All&- &0.01-0.03 & -\\
\hline
  D2& IH & $ \ge 0^{o}$  &$\ge 0.01$ & -\\
 I V& IH+ QD &$\ge 0^{o}$ &0.01-0.03 &- \\
\hline
  D3& NH+ QD &$ >0^{o}$   &$\ge 0.02$ & $<45^{o}$   \\
 M V& All &- & 0.01-0.03& -\\
\hline
 D4& NH+ QD & $ >0^{o}$  &$\ge 0.02$ &$<45^{o}$ \\
 N V& All & -& 0.01-0.03&- \\
\hline
 D5& NH+ QD & $ >2^{o}$  & $\ge 0.01$&- \\
 O V& All &- & 0.01-0.03& \\
\hline
 E1& All & $ >1^{o}$  &$\ge 0.01$ & -\\
 D VI& All & -&0.01-0.03 & \\
\hline
 E2& IH &$\ge 0^{o}$   & $\ge 0.01$ & -\\
 I VI& IH+ QD & $\ge 0^{o}$ &0.01-0.03 &- \\
\hline
 E3& NH+QD &  $>0^{o}$  & $\ge 0.02$  &$>45^{o}$ \\
 J VI& All & - & 0.01-0.03& -\\
\hline
 E4& NH+ QD & $>0^{o}$   & $\ge 0.03$  & $>45^{o}$  \\
 K VI& All & - & 0.01-0.03&- \\
\hline

\end{tabular}
\caption{A comparative study of some predictions obtained in the
present work with earlier analysis\cite{13}. Upper (lower) entry
in each block corresponds to the value obtained in the present
(earlier) analysis.}
\end{center}
\end{table}
\newpage

\begin{figure}
\begin{center}
\epsfig{file=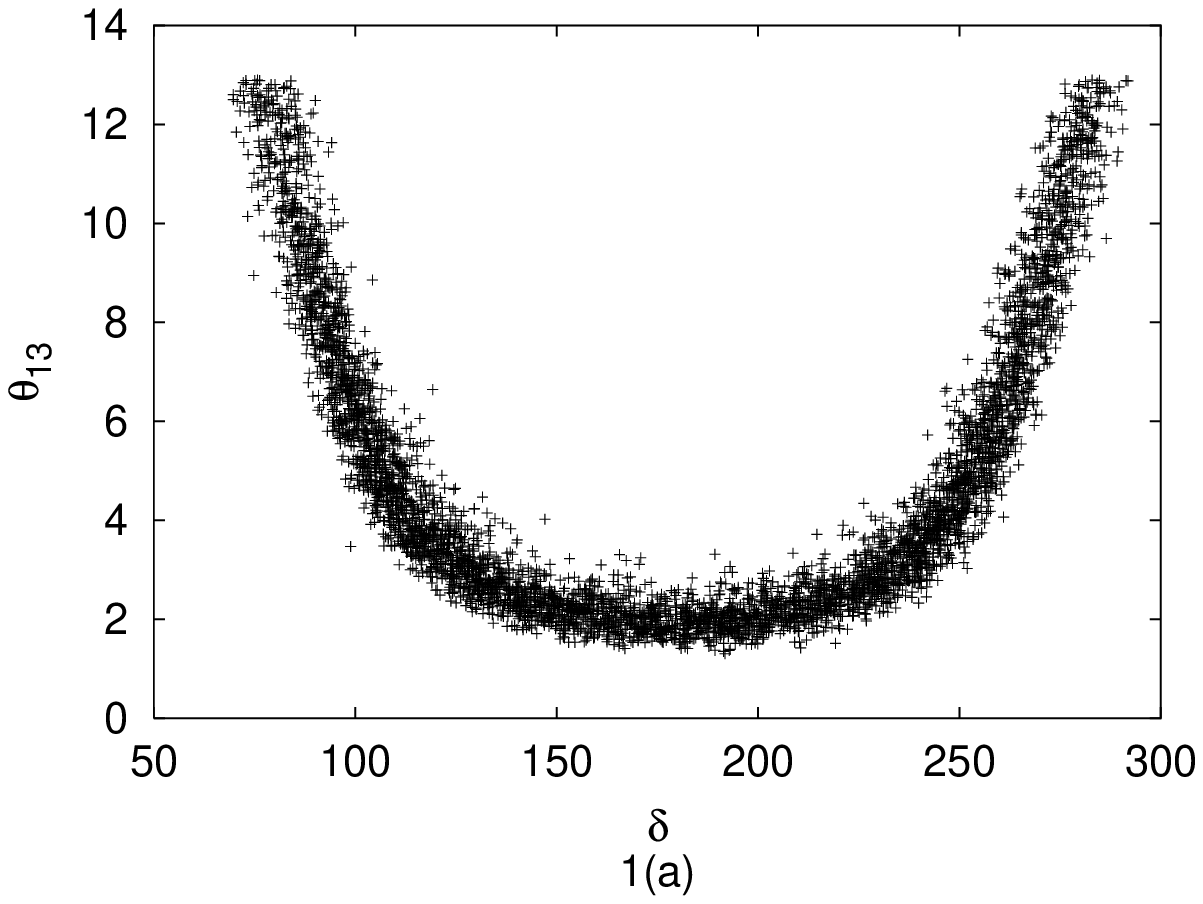,height=5.0cm,width=5.0cm}
\epsfig{file=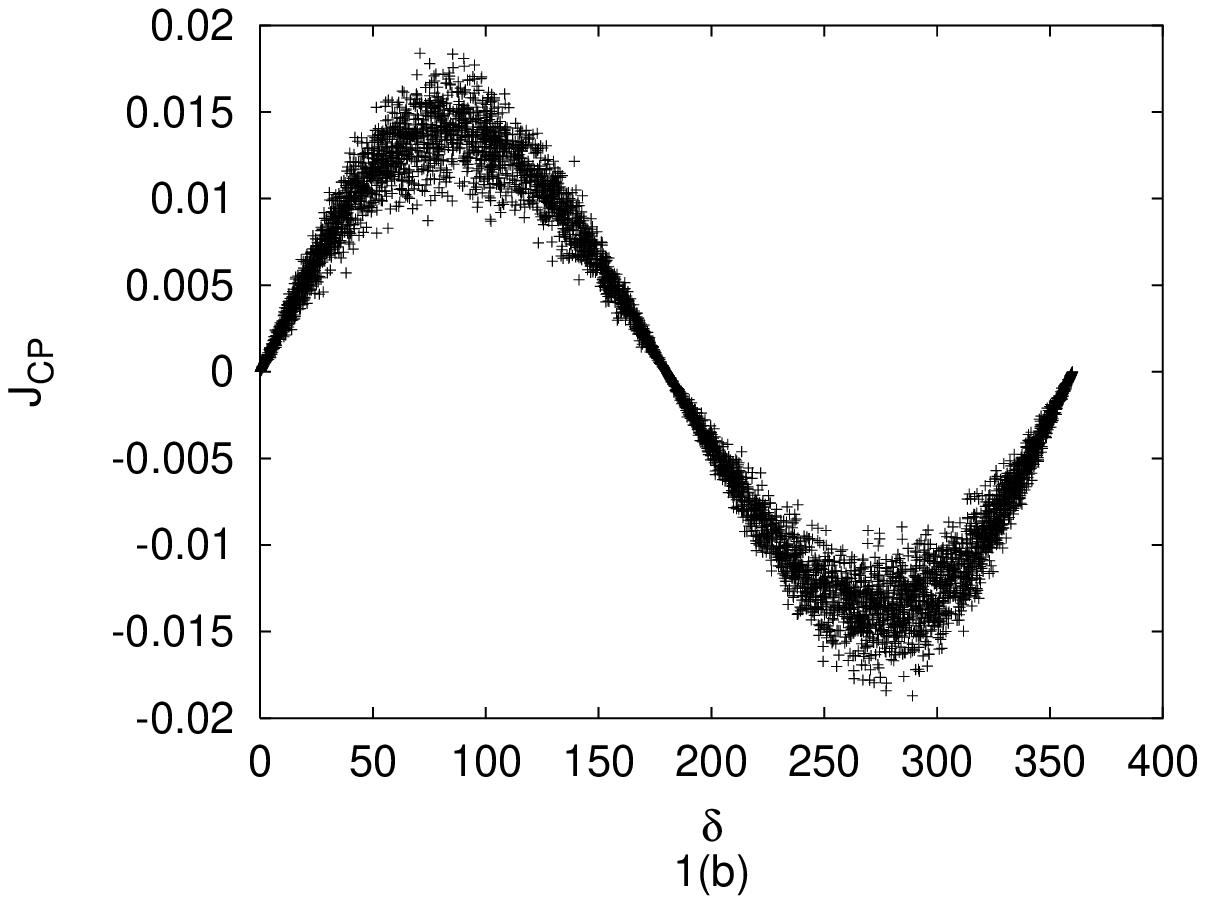,height=5.0cm,width=5.0cm}
\epsfig{file=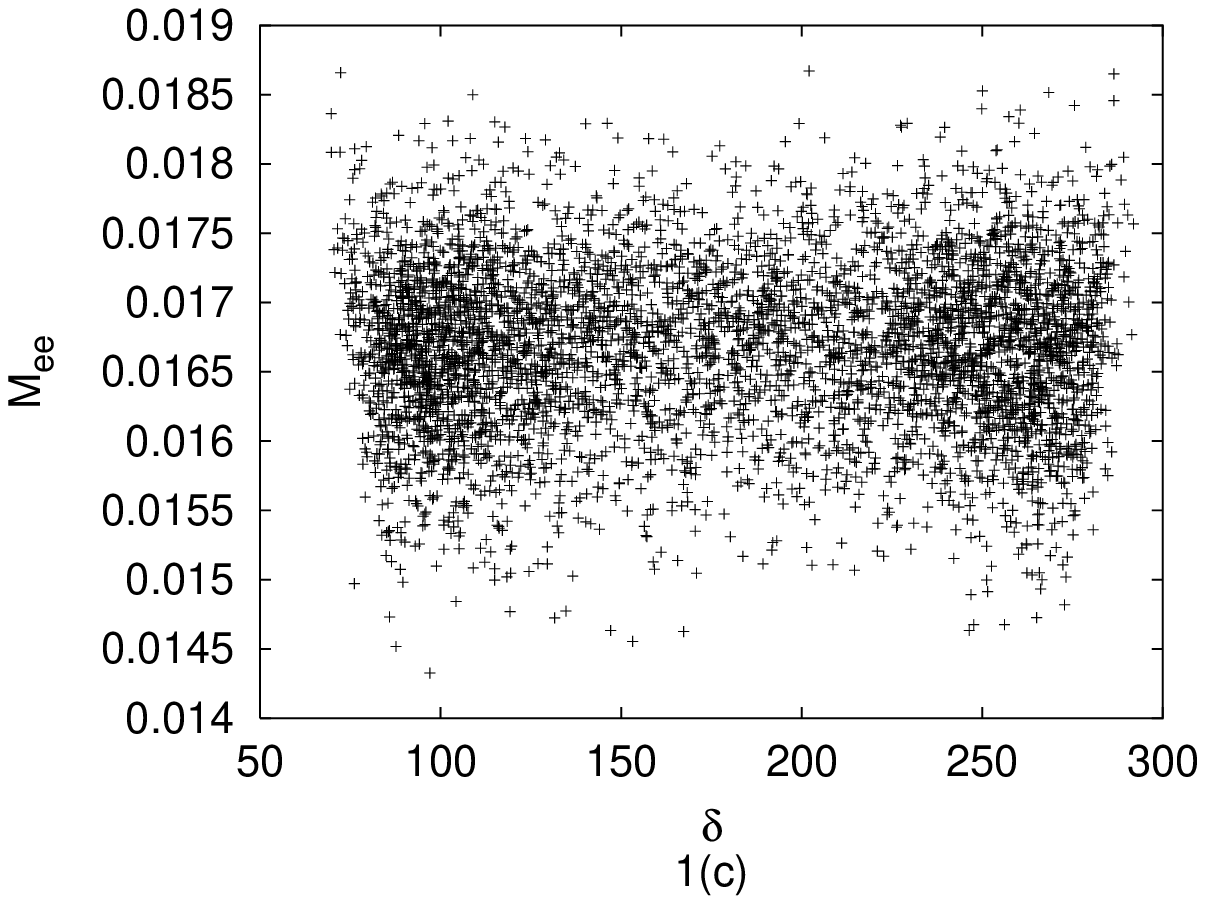,height=5.0cm,width=5.0cm}
\epsfig{file=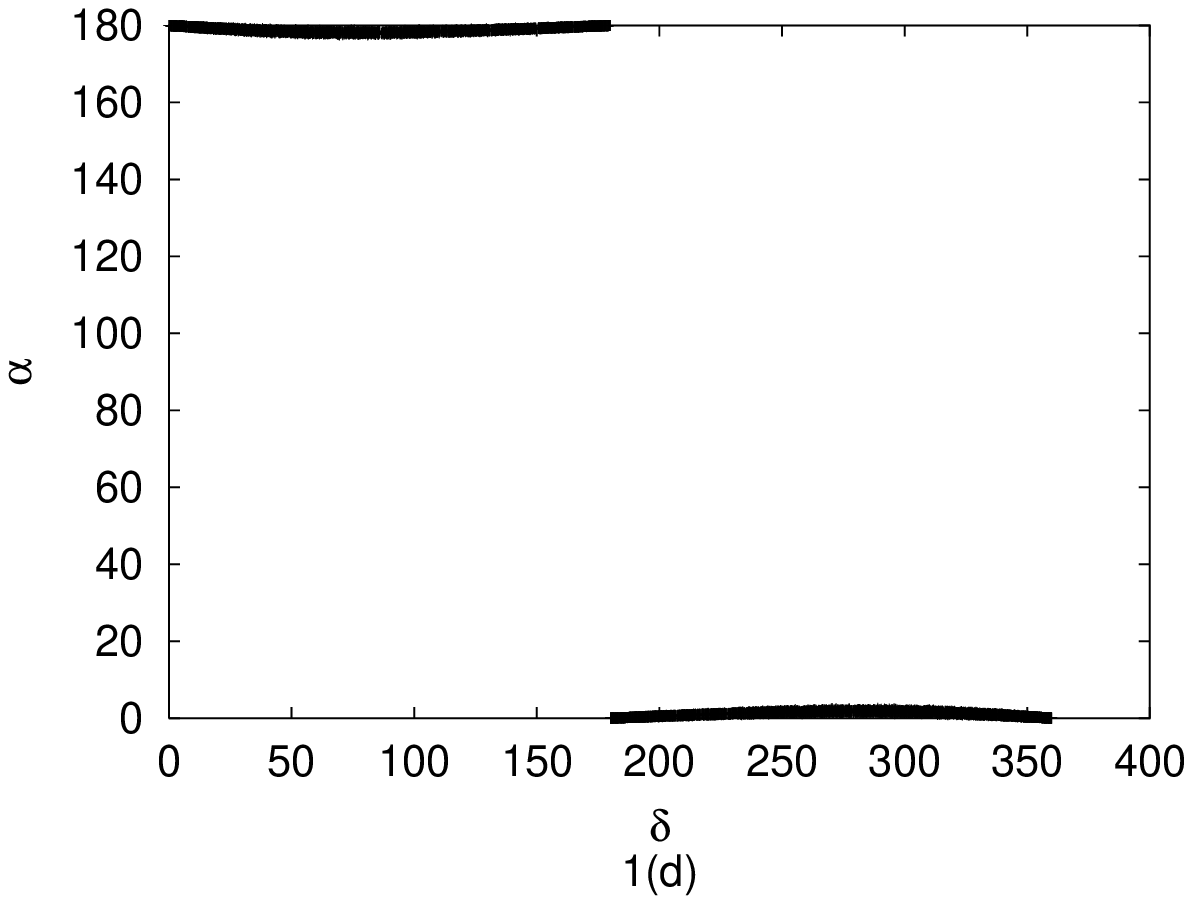,height=5.0cm,width=5.0cm}
\end{center}
\caption{Correlation plots for $A1$ and $B1$ hybrid texture
structures.}
\end{figure}
\begin{figure}
\begin{center}
\epsfig{file=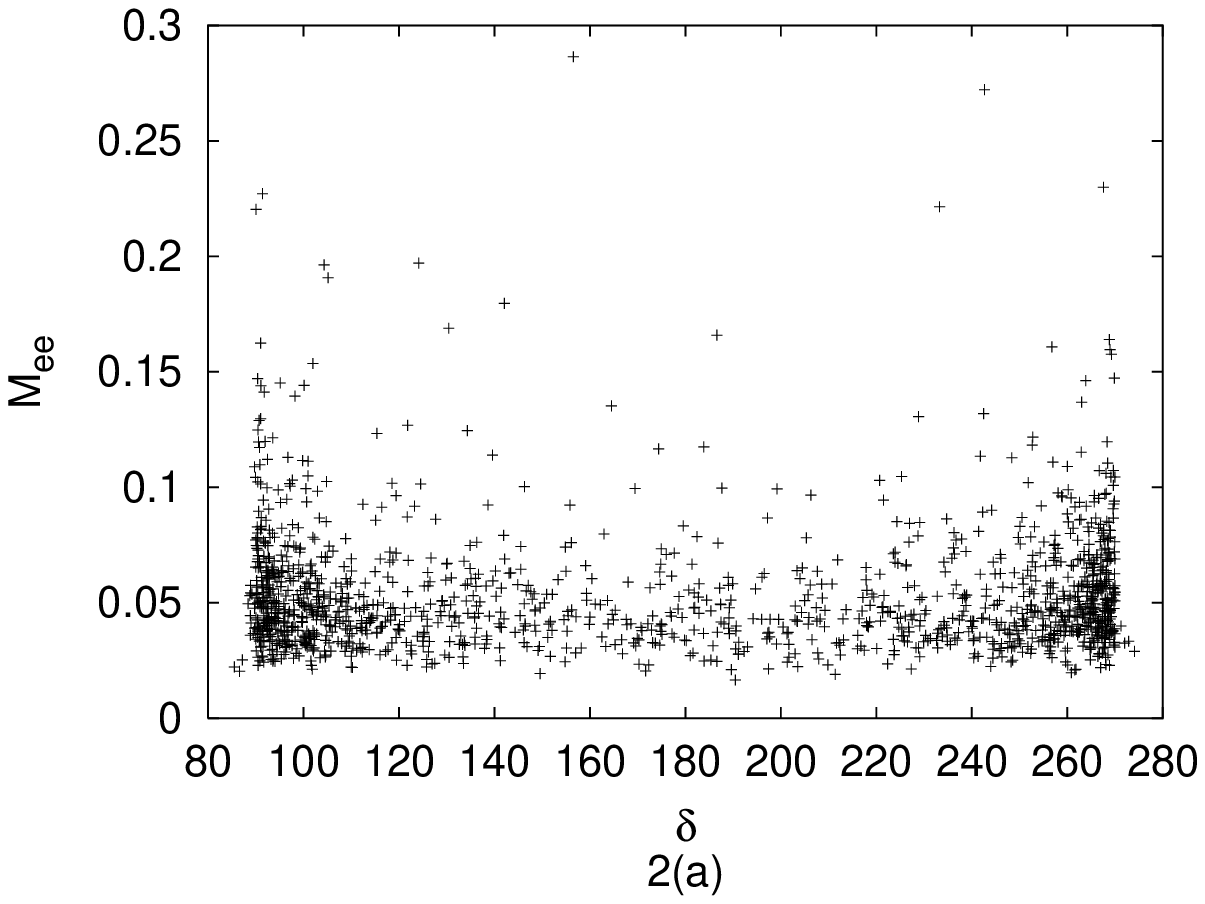,height=5.0cm,width=5.0cm}
\epsfig{file=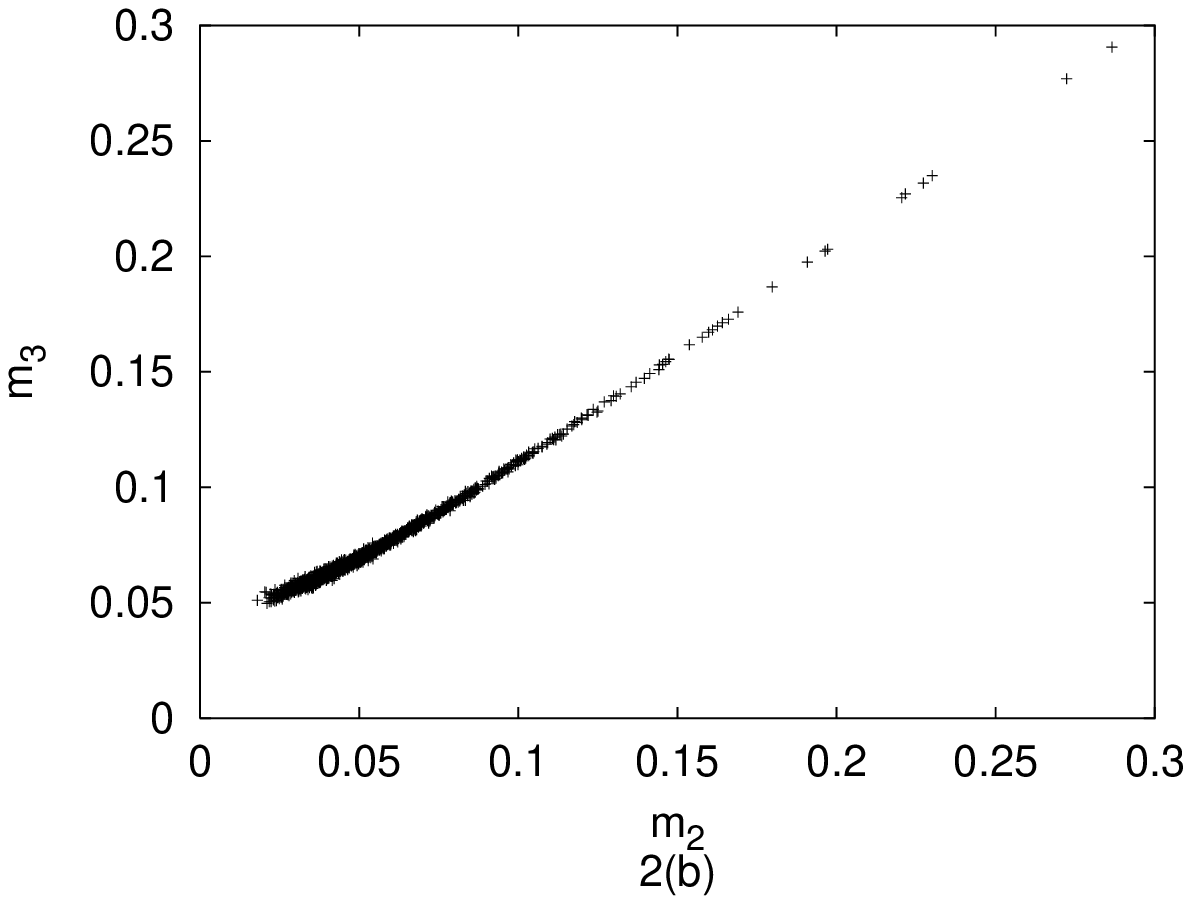,height=5.0cm,width=5.0cm}
\end{center}
\caption{Correlation plots for $B3$ hybrid texture structure.}
\end{figure}
\begin{figure}
\begin{center}
\epsfig{file=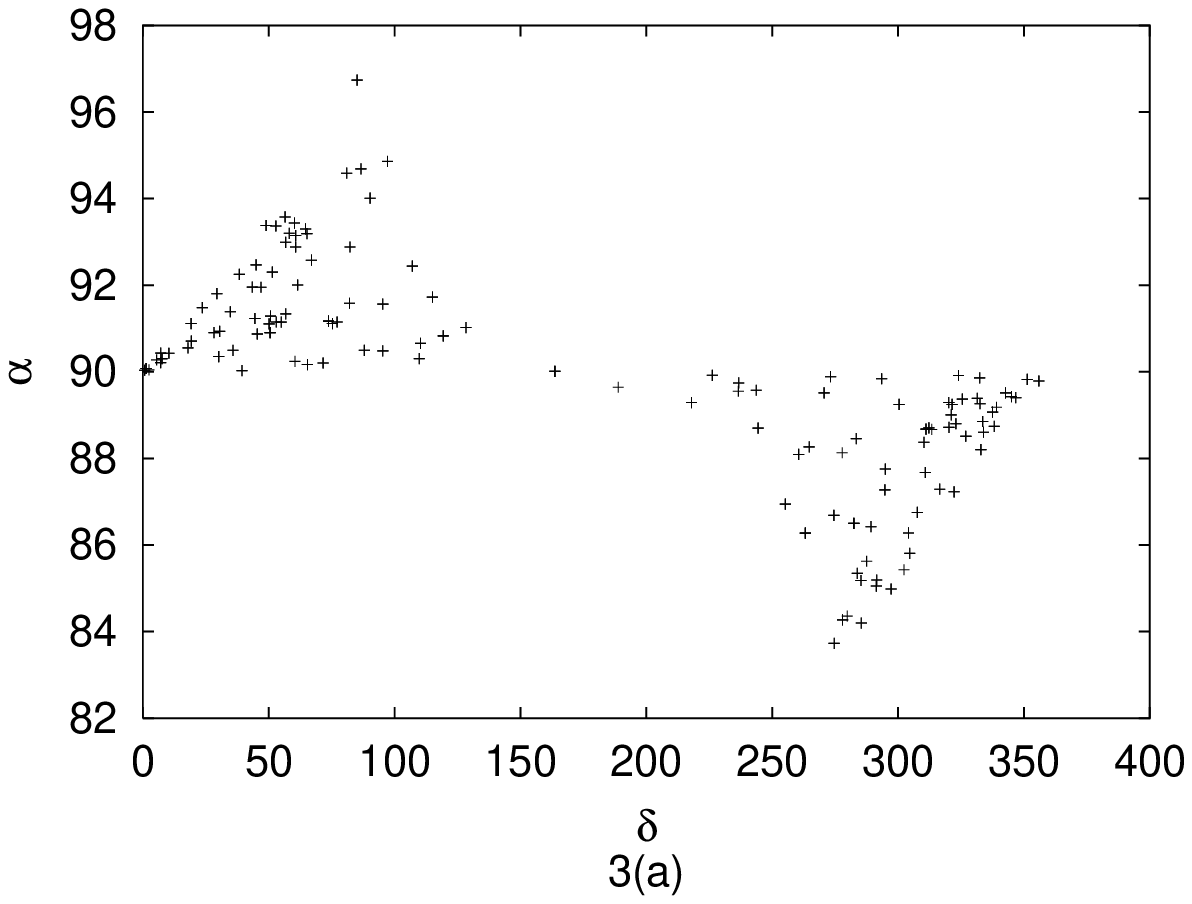,height=5.0cm,width=5.0cm}
\epsfig{file=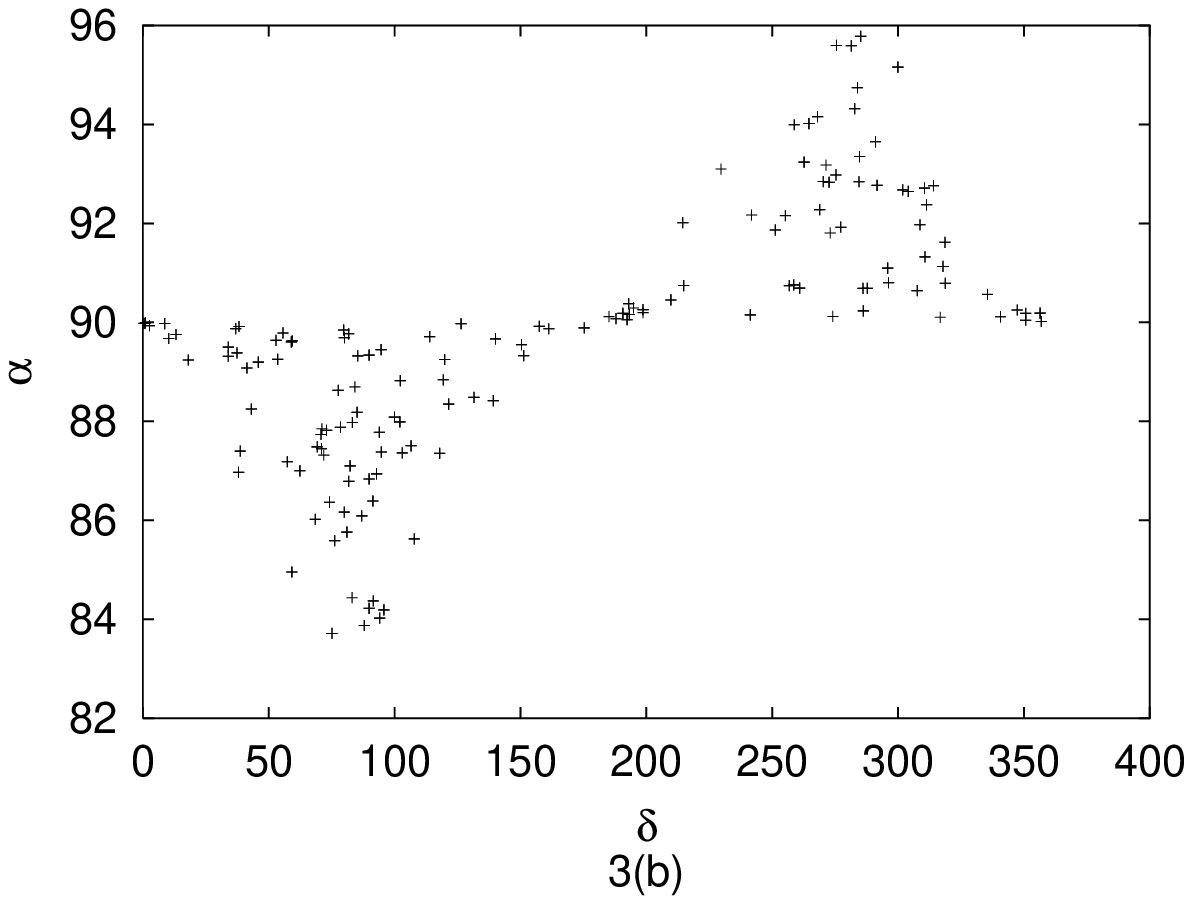,height=5.0cm,width=5.0cm}
\end{center}
\caption{Correlation plots for $D2$ and $E2$ hybrid textures.}
\end{figure}
\begin{figure}
\begin{center}
\epsfig{file=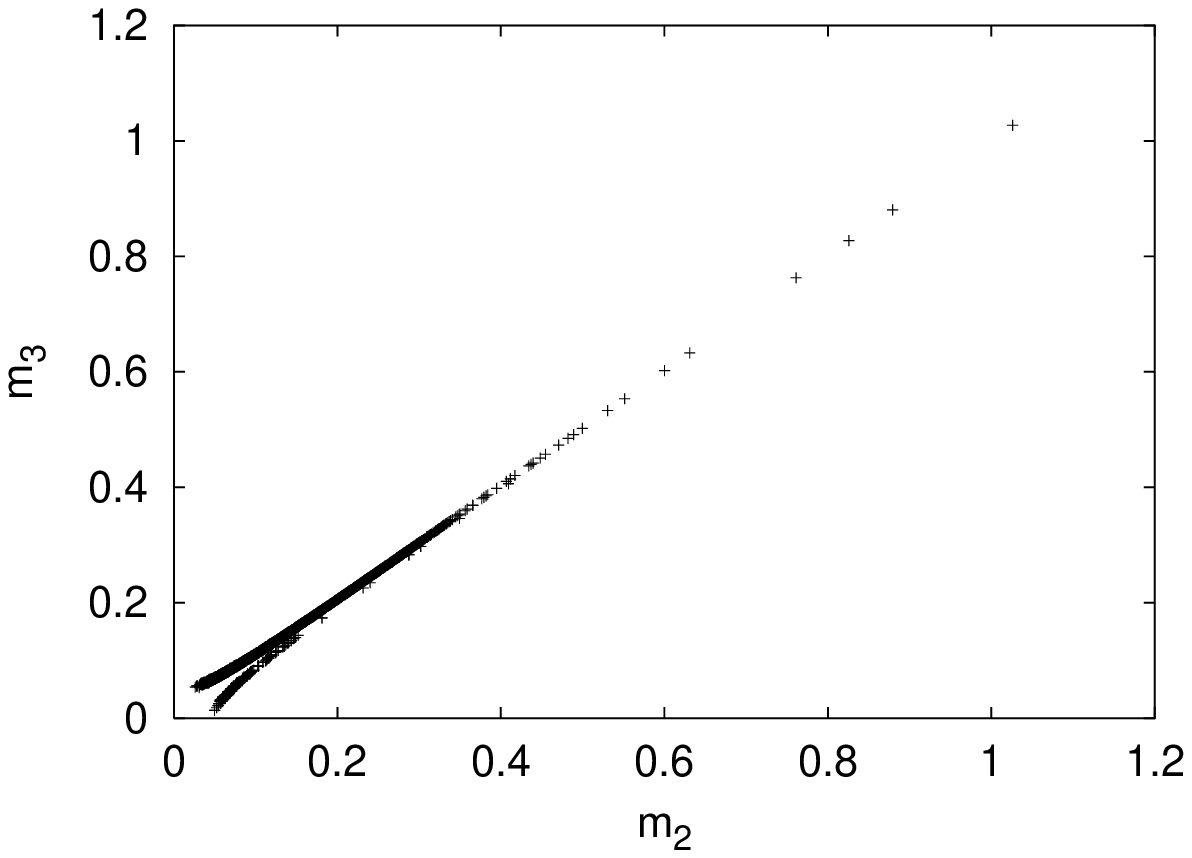,height=5.0cm,width=5.0cm}
\end{center}
\caption{Mass spectrum for hybrid texture $E1$.}
\end{figure}


\begin{thebibliography}{99}





\bibitem{1} H. C. Goh, R. N. Mohapatra and
Siew-Phang Ng, \textit{Phys. Rev.} \textbf{D 68}, 115008 (2003).
\bibitem{2} I. de Medeiros Varzielas , S. F. King and G. G. Ross \textit{Phys. Letts.} \textbf{B 644}, 1532 (2007).
\bibitem{3} M. Frigerio, S. Kaneko, E. Ma and M. Tanimoto \textit{Phys.Rev.}
\textbf{D 71}, 011901 (2005).
\bibitem{4} Paul H. Frampton, Sheldon L. Glashow and Danny
Marfatia, \textit{Phys. Lett.} \textbf{B 536}, 79 (2002); Bipin R.
Desai, D. P. Roy and Alexander R. Vaucher, \textit{Mod. Phys.
Lett} \textbf{A 18}, 1355 (2003).

\bibitem{5}  Zhi-zhong Xing, \textit{Phys. Lett.} \textbf{B 530}
159 (2002).

\bibitem{6} Wanlei Guo and Zhi-zhong Xing, \textit{Phys. Rev.} \textbf{D 67},
053002 (2003).

\bibitem{7} Alexander Merle and Werner Rodejohann, \textit{Phys. Rev.}
\textbf{D 73}, 073012 (2006); S. Dev and Sanjeev Kumar,
\textit{Mod. Phys. Lett.} \textbf{A 22}, 1401(2007), arXiv:0607048
[hep-ph].

\bibitem{8} S. Dev, Sanjeev Kumar, Surender Verma and Shivani Gupta,
 \textit{Nucl. Phys.} \textbf{B 784}, 103-117 (2007).
\bibitem{9} S. Dev, Sanjeev Kumar, Surender Verma and Shivani Gupta,
 \textit{Phys. Rev.} \textbf{D 76}, 013002 (2007).
\bibitem{10} M. Honda, S. Kaneko and M. Tanimoto, \textit{JHEP}
\textbf{0309}, 028(2003).
\bibitem{11} S. Kaneko, M. Tanimoto, \textit{Phys. Lett.} \textbf{B 551}, 127 (2003).
\bibitem{12}M. Frigerio, A. Y. Smirnov, \textit{Phys. Rev.}
\textbf{D 67}, 013007 (2003).

\bibitem{13} S. Kaneko, H.Sawanaka and M. Tanimoto, \textit{JHEP}
\textbf{0508}, 073(2005).


\bibitem{14} Srubabati Goswami, Subrata Khan, Atsushi Watanabe, arXiv:0811.4744
v1 [hep-ph], Amol Dighe, Narendra Sahu, arXiv:0812.0695v2 [hep-ph]
\bibitem{15} G. L. Fogli \textit{et al}, arXiv:0506083 [hep-ph].

\bibitem{16} M. C. Gonzalez-Garcia, Michele Maltoni, \textit{Phys. Rept.} 460 (2008) 1-129, arXiv:0704.1800v2 [hep-ph].

\bibitem{17} C. Jarlskog, \textit{Phys. Rev. Lett.} \textbf{55}, 1039 (1985) .
\bibitem{18}http://doublechooz.in2p3.fr/; F. Ardellier \textit{et al.} (Double
Chooz Collaboration), arXiv:0606025 [hep-ex].
\bibitem{19} http://dayabay.ihep.ac.in/; Daya Bay Collaboration, arXiv:0701029 [hep-ex].
\bibitem{20} H. V. Klapdor- Kleingrothaus, \textit{Nucl. Phys.Proc.Suppl.}\textbf{145}, 219 (2005).
\bibitem{21} Arnaboldi C \textit{et al.} 2004a \textit{Nucl. Instrum. Meth.}\textbf{A 518} 775
\bibitem{22} Arnaboldi C \textit{et al.} (CUORICINO collaboration) \textit{Phys. Lett. B} \textbf{584} 20 (2004)
\bibitem{23} I. Abt \textit{et al.}(GERDA collaboration) arXiv:0404039 [hep-ex].
\bibitem{24} Sarazin X \textit{et al.} 2000 Preprint arXiv:0006031
[hep-ex].
\bibitem{25} Hisakazu Minakata, arXiv:0701070 [hep-ph].
\end{thebibliography}
\end{document}